\documentclass[aps,superscriptaddress,prl,reprint, showpacs
,amsmath,amssymb,twocolumn]{revtex4-1}

\usepackage{graphicx}
\usepackage[utf8]{inputenc}

\usepackage{amsmath}
\DeclareMathOperator\erf{Erf}

\begin{document}

\title{A complete mean-field theory for dynamics of binary recurrent networks}

\author{Farzad Farkhooi}
\author{Wilhelm Stannat} 

\affiliation{Institut f\"ur Mathematik, Technische Universit\"at Berlin, 10623 Berlin,Germany.} 
\affiliation{Bernstein Center for Computational Neuroscience, 10115 Berlin, Germany. }


\begin{abstract} 
We develop a unified theory that encompasses the macroscopic dynamics of
	recurrent interactions of binary units within arbitrary network
	architectures. 
	Using the martingale theory, our mathematical analysis provides a
	complete description of nonequilibrium fluctuations in networks with
	finite size and finite degree of interactions. Our approach allows the
	investigation of systems for which a deterministic mean-field theory
	breaks down.
	To demonstrate this, we uncover a novel dynamic state in which a
	recurrent network of binary units with statistically inhomogeneous
	interactions, along with an asynchronous behavior, also exhibits
	collective nontrivial stochastic fluctuations in the thermodynamical
limit.  
\end{abstract}

\maketitle

An important means to understand the collective dynamics of high-dimensional
spin systems is to use mean-field theory (MFT) to describe the activity of the
system population in terms of associated lower dimensional dynamics
\cite{mezard_spin_1986}.  The classical characterization of the emerging states
uses the analysis of the systems' Hamiltonian functions \cite{ising_beitrag_1925,
*hohenberg_theory_1977}.  However, this powerful approach cannot be applied, if
the underlying interactions among units are directed and asymmetrically
disordered as in various soft materials \cite{hamley_introduction_2013} and in
particular recurrent neuronal networks \cite{van_vreeswijk_chaotic_1998},
in contrast to the  bidirectionality of interactions in spin glasses \cite{
mezard_spin_1986}. The
formulation of MFTs in these cases typically assumes
statistically homogeneous interactions among units in extensively large
networks \cite{van_vreeswijk_chaotic_1998, glauber_timedependent_1963}.  
The standard theoretical framework here is to expand the system's master
equation (so-called Kramers-Moyal expansion), in which the lowest order tree
level expansion of these theories yields the mean-field limit and systematic
corrections can be obtained by perturbative and renormalization group methods
\cite{hohenberg_theory_1977}.  In the case of statistically homogeneous binary
recurrent networks various methods have been used to obtain
finite-size fluctuations
\cite{ginzburg_theory_1994,*buice_systematic_2010,*renart_asynchronous_2010}
and an interesting approach to analyze the mean-field limit of a single instance
of asymmetric Ising networks has been investigated \cite{mezard_exact_2011}.
However, no general theory has been developed to treat systems with
statistically inhomogeneous and asymmetric interactions.
In this letter, we use a surprisingly elementary method that can be used to
remove the need for these assumptions by deriving a novel MFT
that captures the dynamic behavior of recurrent networks with binary units,
including finite-size effects on population fluctuations.  In this framework,
we isolate the finite-size fluctuation of the system in the martingale
structure of the network's Markovian dynamics and derive the macroscopic
behavior of the system given the gain function of individual units.  Our
mathematical approach readily identifies the conditions on the connectivity
structure that are necessary to guarantee the convergence of the average
population activity to a deterministic limit.  Furthermore, our analysis
reveals a novel dynamic state in a network with inhomogeneous coupling,  in
which the large amplitude fluctuations of the average population activity
survive irrespective of the network size.  Such stochastic synchronization
could be relevant for the description of collective neocortical network
dynamics.

Consider a {\it model network} that is described by an adjacency binary matrix
$\mathbf{J}=(J_{ij})$ of $N$ binary units, whose current states are
denoted as $\mathbf{n}(t) := (n_1(t),\ldots, n_N(t))$. The vector
$\mathbf{n}(t)$ is a time-continuous Markov chain on $\{0,1\}^N$ with
rate matrix, where $Q(\mathbf{n},\mathbf{m}) = 0$ if and only if
$||\mathbf{n}-\mathbf{m}||\ge 2$ and 
$$ Q(\mathbf{n},\mathbf{m}) = \begin{cases} f_i (\mathbf{n}) & \mbox{ if }
	\mathbf{m}-\mathbf{n} = \mathbf{e_i} \\ 1-f_i (\mathbf{n}) & \mbox{ if }
\mathbf{m}-\mathbf{n} = - \mathbf{e_i} \end{cases} $$ 
where $e^{(i)}_j = \delta_{ij}$ denotes the $i$-th unit vector. In order to
comply with the centralization property of $Q$-matrices, it follows that: 
$ Q(\mathbf{n},\mathbf{n}) = - \sum_{i=1}^{N} n_i(t)\big(1 -
2f_i(\mathbf{n}(t))\big)+f_i(\mathbf{n}(t)) .$
The analytical gain function $f_i(\mathbf{n}(t))$ defines the state transition
rate of a unit $i$, given the state of the network, $\mathbf{n}(t)$, and it is
assumed to take values in the range $[0,1]$.  Typically, this function is
written as $f_i(u_i(t))$, where $u_i(t)$, which represents the input to the
unit $i$ with the scaling parameter $0<\gamma$, is written as
\begin{equation} u_i(t) := \bar{J}K_i^{-\gamma} \sum_{j=1}^N J_{ij}n_j(t) +
K_i^{1-\gamma} \mu_{0,i} \label{input_eq} \end{equation} 
where $\bar{J}$, $K_i$, and $\mu_{0,i}$ are the coupling strength, the number
of recurrent input units ($K_i := \sum_j^N J_{ij}$) and the external drive to
the $i$-th unit, respectively.  For the convenience of the current
presentation, we consider here networks with $K_i=K$, $f_i=f$ and $\mu_{0,i}=
\mu_0$ for all $i$. We will provide below (in eqns.~\ref{CI} and \ref{CII})
conditions on the network structure, $\mathbf{J}$, that imply the convergence
of the averaged population activity in the network towards a deterministic
limit 
\begin{equation} m(t) \stackrel{!}{=} \lim_{N\rightarrow \infty}
\frac{1}{N}\sum_{i=1}^N n_i(t).  \label{LLN} \end{equation}
Here, $m(t)$ is known as the mean-field limit and has the following temporal
dynamics: 
\begin{equation} \frac{d}{dt} m(t) = -m(t) + F(m(t)) \label{MFE} \end{equation}
for some a priori unknown function $F$. In order to determine $F$, we use the
following semimartingale decomposition, that specifies the difference between
$\bar{n}(t):=\frac{1}{N}\sum_{i=1}^N n_i(t)$ (i.e., the average population
activity of a finite-size network) and the mean-field limit $m(t)$ of the
system,
\begin{align} & \bar{n}(t)  - m(t)  = (\bar{n}(0) - m(0)) - \int_0^t ds \;
\big( \bar{n}(s) - m(s)\big)  \nonumber \\ & \quad + \int_0^t ds \; \big(
\frac{1}{N}\sum_{i=1}^N f(u_i(t)) - F(m(s)\big) + \mathcal{M}(t) , \label{smar}
\end{align}
where $\mathcal{M}(t)$ is some square integrable martingale that, according to
the general theory of Markov processes \cite{stroock_introduction_2008},
satisfies
\begin{equation} \operatorname{E} \left( \mathcal{M}(t)^2 \right) =
\frac{1}{N^2} \int_0^t ds \; \operatorname{E}(-Q(\mathbf{n}(s),\mathbf{n}(s)))
\leq \frac{t}{N}.  \label{mdec} \end{equation}
Note that $\operatorname{E}\left[\mathcal{M}(t)\right]=0$ and, in general,
$\mathcal{M}(t)$ specify finite-size fluctuations in the average population
activity. Provided that $m(t)$ exists (refer to eqns.~\ref{CI} and \ref{CII}
for a justification of this ansatz), we can construct the function $F$ by
expanding $\frac{1}{N}\sum_{i=1}^N f(u_i(t))$ as $N\rightarrow\infty$ in
eqn.~\ref{smar} around
\begin{equation} \mu_1(t) := K^{1-\gamma}\big(\bar{J}m(t) + \mu_{0}\big)\, .
\label{mu_1} \end{equation}
Using the lemma that is described in 
\cite{[{See Supplementary Matrial [url], which includes Refs. \cite{
gillespie_general_1976, doob_topics_1942}}] supp}
, we obtain the
following series expansion  
\begin{equation} F(m(t)) = f(\mu_1(t)) + \sum_{r=2}^{\infty}
\frac{f^{(r)}(\mu_1(t))}{r!}\mu_{r}(t).  \label{F} \end{equation}
where $\mu_1$ represents the average input to a unit in the network at time
$t$. The higher order coefficients can be computed by expanding
$\mu_r:=\lim_{N\rightarrow\infty}\frac{1}{N}\sum_{i=1}^N [(u_i - \mu_1)^r]$. 
The second order coefficient is given by:
\begin{equation} \mu_2(t) = \bar{J}^2K^{1-2\gamma}m(t)(1-m(t))  \label{mu_2}
\end{equation}
and the subsequent coefficients are given by: 
\begin{equation} \mu_r(t) = \bar{J}^r K^{-r\gamma}\sum_{q=0}^{r} a_q m(t)^q
\sum_{s=0}^{r-q} b_s m(t)^s \label{mu} \end{equation}
where, 
$$ a_q := {r \choose q} (-1)^q K^{q} $$ and $$ b_s :=\mathcal{S}(r-q,s)(K)_s \,
$$ 
Here, $\mathcal{S}$ is a Stirling number of the second kind and $(.)_s$ denotes
the falling factorial. In the binomial expansion of $\mu_r(t)$ given in
eqn.~\ref{mu}, the summation over $j$ (note that $j$ is hidden in the
definition of $u_i$) is performed using the ansatz that $m(t)$ exists;
thereafter, summation over $i$ in  the average operator
$\lim_{N\rightarrow\infty}\frac{1}{N}\sum_{i=1}^{N}[.]$ is applied.  In order
to provide the sufficient conditions for the existence of a deterministic
limit, $m(t)$, the summation order must be changed.  Therefore, the first
condition for $m(t)$ and $\mu_1(t)$ to exist is
\begin{equation} 	\lim_{N \rightarrow\infty} \frac{1}{N^2} \sum_{j=1}^{N}
\left( \sum_{i=1}^{N}  \left( J_{ij} - \frac{K}{N} \right) \right)^2 = 0 .
\label{CI} \end{equation}
This condition essentially states that column sums distribution of the
connectivity matrix must obey the weak Law of Large Numbers (LLNs) and
eqn.~\ref{CI} implies that the coefficient in front of $f^\prime$ in the series
expansion that leads to eqn.~\ref{F} vanishes in the thermodynamic limit
\cite{supp}.  The second condition for the pointwise convergence of an averaged
population activity to the MFT in eqn.~\ref{LLN} is given by 
\begin{equation} \lim_{N \rightarrow \infty} \frac{1}{N^2}  \sum_{j_1 \neq
j_2}^N \left( \sum_{i=1}^N \left( J_{ij_1}J_{ij_2} -
\frac{K(K-1)}{N(N-1)}\right) \right)^2 = 0.  \label{CII} \end{equation}
This condition specifies that, as $N \rightarrow \infty$, the mean covariance
of columns in the connectivity matrix $\mathbf{J}$ must satisfy the LLNs.  The
higher order condition can be similarly determined in order to achieve a
pointwise convergence of the averaged population activity to its mean-field
limit, as described  in \cite{supp}.
An important result here is that the condition in eqn.~\ref{CI} implies that
eqn.~\ref{CII} and all higher order conditions are satisfied for all
fixed-in-degree networks and {\it iid} connectivity matrices and therefore the
MFT in eqn.~\ref{MFE} becomes universal for those coupling structures.


In the above calculation, we assume networks with a finite input connections
per unit (i.e., $K$). However, it is often of interest to study network
dynamics when the number of inputs into units is large (e.g. $K\rightarrow
\infty$).  In order to study this classical case, we must investigate the
asymptotic behavior of $\mu_r$ in eqn.~\ref{mu} in the order of $K$; it can be
observed that the odd coefficients are given by
$$ \mu_{2k+1} \sim \mathcal{O}(K^{1-(2k+1)\gamma}) $$ 
and the even coefficients are given by
$$ \mu_{2k} \sim \mathcal{O}(K^{1-2k\gamma}) + (2k-1)!! \; \mu_2^k , $$
for $k \in \mathbb{N}$.  Hence, it is apparent that the scaling parameter
$\gamma$ plays a critical role in the large $K$ limit.  The scaling parameter
$\gamma$ is generally assumed to take the value $0.5$; in this case, $\mu_2\sim
\mathcal{O}(1)$ and the mean-field coefficients of eqn.~\ref{MFE} converge as
$K\rightarrow \infty$, towards the central moments of a Gaussian distribution
function and the network can be asynchronous similar to the nonequilibrium and
chaotic dynamics observed in \cite{van_vreeswijk_chaotic_1998}. As a result, the
related power series that is given by eqn.~\ref{F} can be reformulated in terms
of a simple Gaussian integral; in this special case, eqn.~\ref{MFE}
reduces to
\begin{equation} \frac{d}{dt} m (t) = -m (t) + \int dx \; f(x)\mathcal{N}(x;
\mu_1, \mu_2) ,\label{MFEI}  \end{equation}
where $\mathcal{N}$ is a Gaussian density. In the above analysis, we first take
$N\rightarrow\infty$ to arrive at the mean-field of eqn.~\ref{MFE} and then we
consider $K\rightarrow \infty$ in order to recover eqn.~\ref{MFEI}.  This
derivation recovers the result has been previously known
\cite{dahmen_correlated_2016, van_vreeswijk_chaotic_1998}, while provides
insight on the structure of corrections to Gaussian density for finite $K$
networks.
Our analysis here shows that the finite $K$ correction to eqn.~\ref{MFEI} is
relatively small.  Thus, using asymptotic corrections up to the $\theta$-th
order to the Gaussian density, the function $F$ for a finite $K$ is given by 
\begin{equation} F(m(t))= \int dx \,f(x)
(1+\mathcal{G}_{\theta}(x))\mathcal{N}(x;\mu_1,\mu_2), \label{gram_exp}
\end{equation} 
where, $\mathcal{G}_{\theta}(x):= \sum_{k=3}^{\theta} \frac{(-1)^k \mu_k}{k!
\, \mu_2^{k/2}} H_k(\frac{x-\mu_1}{\sqrt{\mu_2}}) $; here, $H_k$ is a Hermite
polynomial of $k$-th order.  This representation is the usual form of the
Gram-Charlier expansion (the so-called Type A series) is an expansion of a
probability density function about a Gaussian distribution with common $\mu_1$
and $\mu_2$ \cite{stuard_kendalls_1994}.
This expansion has been used in eqn. C2 of Dahmen et al.
\cite{dahmen_correlated_2016} to include finite-size corrections due to
pair-wise correlations in the MFT.
The structure of centralized moments in eqn.~\ref{mu} allows for an arbitrary
precise calculation of the mean-field limit.
It is noteworthy that eqn.~\ref{gram_exp} is the steady-state mean-field limit
for all possible fixed-indegree networks.


The semimartingale decomposition that is given in eqn.~\ref{smar} provides
 information on the {\it finite size} scaling of the system.
Using eqn.  \ref{mdec}, we can determine the fluctuations magnitude of the
average population activity in finite networks in the mean-square sense as  
$$ \operatorname{E} \left( \mathcal{M}(t)^2 \right) = \frac{1}{N^2} \int ds
\operatorname{E}\big(\sum_{i=1}^N  n_i(s)(1-2f(u_i(s)))+f(u_i(s))\big)  $$
and, by expanding $\frac{1}{N} \sum_{i=1}^N f(u_i(t))$ at $\mu_1 (t)$, we
arrive at 
\begin{equation} \operatorname{E} \left( \mathcal{M}(t)^2 \right) = \frac{1}{N} \int_0^t ds
\big(m(s)(1-2(g(\mu_1)+\mathcal{R})) + g(\mu_1)+ \mathcal{R})\big) 
	\label{m_msq}
\end{equation}
where $g(\mu_1):= f(\mu_1(t)) + f''(\mu_1)\mu_2/2$ and $\mathcal{R} :=
\sum_{r=3}^{\infty} \frac{f^{(r)}(\mu_1)}{r!}\mu_r$ denotes the remainder terms
of the expansion. The average population activity dynamics of a finite size
network can be described approximately in terms of the following
Ornstein-Uhlenbeck process 
\begin{equation} d \bar{n}(t) \approx \big(- m(t) + F(m(t))\big)dt +
\frac{\sigma(t)}{\sqrt{N}}d\mathcal{B}_t \label{OU_rate} \end{equation}
where $\sigma^2(t) := m(t)(1-2g(\mu_1(t))+g(\mu_1(t))$ and $\mathcal{B}_\cdot$
is Brownian motion. In the approximation of eqn.~\ref{OU_rate}, we ignore the
contribution of remainder terms (e.g., $\mathcal{R}$) to $\sigma (t)$.  Our
result recovers previously known scaling of the finite size fluctuations \cite{
	brunel_fast_1999, *mattia_mean-field_2002, *grytskyy_unified_2013}
using the semimartingale method.


\begin{figure}
	\centering
	\includegraphics{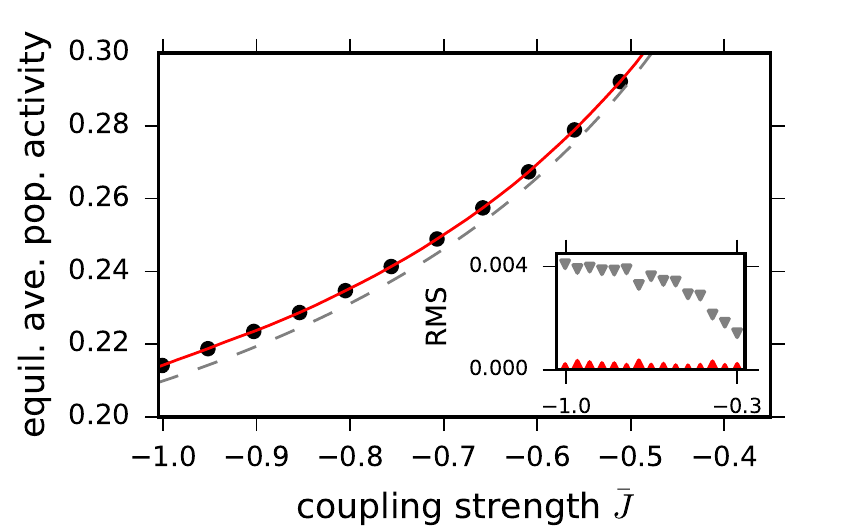}
	\caption{Convergence of the average population activity to the
	steady-state MFT predictions. The \textbf{red line} indicates the
	predictions of complete MFT up to fifth order correction. The
	\textbf{dashed gray line} is the predictions of mean-field eqn.~\ref{MFEI},
	assuming only Gaussian fluctuations. The \textbf{black dots} represent
	network simulations averaged over 20
	independent trails (error bars are smaller than symbol size). The
	\textbf{inset} is the Root Mean Square of error (RMS) between simulations and
	the complete theory (\textbf{upward red triangles}) and the Gaussian
	approximate theory (\textbf{downward gray triangles}). The simulations
	were performed using a Gillespie algorithm for $T= 5\times10^5$
	steps with the gain function given by eqn.~\ref{gain} . The averaged
	activity was estimated in the last $5\times10^3$ steps across all
	trials.  Parameters: $N=1000$, $\gamma=0.5$, $\alpha=5$, $K=10$ and $\mu_0=0.1$.
} \label{fig1} \end{figure}

In order to {\it demonstrate the applicability} of our approach, we consider two
scenarios that are relevant to the theoretical analysis of neural systems. The
first scenario is that units receive a constant external input $\mu_0>0$. When
$\bar{J} < 0$ and $\gamma=0.5$ this system exhibits a nonequilibrium and
chaotic state for which the external input is canceled by internal recurrent
dynamics \cite{van_vreeswijk_chaotic_1998}.  We choose a widely-used gain
function in neural networks theory which it is given by
\begin{equation} f(x):=\frac{1+\erf(\alpha x)}{2}.  \label{gain} \end{equation}
The parameter $\alpha$ describes the intrinsic noise intensity of the
individual units and therefore must be positive. When
$\alpha\rightarrow\infty$, this transfer function approximates to the
well-studied Heaviside step function
\cite{van_vreeswijk_chaotic_1998,dahmen_correlated_2016}.  Using the transfer
function given by eqn.~\ref{gain} (with $\alpha = 5$) and a directed
fixed-in-degree Erd\H{o}s-R\'{e}nyi network (with $K=10$), we compute the
complete steady-state mean-field limit using eqn.~\ref{gram_exp} by including
up to the fifth order corrections (Fig.\ref{fig1}, red line). We compare the
complete MFT (Fig.\ref{fig1}, red line) with the mean-field prediction that
assumes only Gaussian statistics (i.e., $K\rightarrow\infty$) in
eqn.~\ref{MFEI} (Fig.\ref{fig1}, dashed gray line). The difference between the
predictions becomes apparent as $|\bar{J}|$ increases.  Numerical simulations
of a finite-size network ($N=1000$) are used to estimate the steady-state
population activity by averaging 20 independent trials (Fig.\ref{fig1}, black
dots).  The equilibrium population average activity of simulated networks
(Fig.\ref{fig1}, black dots) exhibits excellent agreement with both the
complete (Fig.\ref{fig1}, red line) and the Gaussian approximation
(Fig.\ref{fig1}, dashed gray line) of the mean-field limit in the case of weak
coupling.  However, in cases where the coupling is strong, the average
population equilibrium activity deviates from the Gaussian approximation
(Fig.\ref{fig1}, dashed gray line) and, instead, follows the predictions of the
complete mean-field limit (Fig.\ref{fig1}, red line).  Therefore, the Gaussian
approximation that is given in eqn.~\ref{MFEI} is only reasonable for weak
coupling and relatively large value of $K$.  The error between steady-state
population activity from the simulations (Fig.\ref{fig1}, black dots) and the
Gaussian approximation (Fig.\ref{fig1}, dashed gray line) increases as
$|\bar{J}|$ becomes larger (Fig.\ref{fig1}, downward gray triangles in the
inset), in contrast to the complete MFT stays constant (Fig.\ref{fig1}, upward
red triangles in the inset).

\begin{figure} \centering \includegraphics{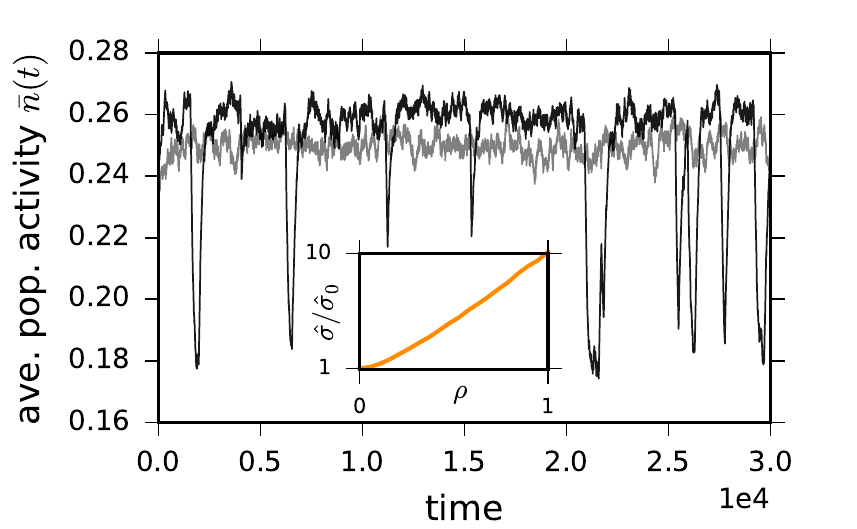} \caption{Emergence of
	stochastic MFT. The \textbf{black line} shows the temporal evolution of
	a simulated network; this network does not
	have a deterministic MFT since the condition in eqn.~\ref{CI} is not
	satisfied.  The out-degree of a single unit in the network was set to
	be $N$ (i.e.,  $\rho=1$ in eqn.~\ref{F_s}). For comparison, the
	\textbf{gray line} shows the average population activity of a similar
	network for which $\rho=K/N$. 
	The \textbf{inset} indicates the normalized empirical
	standard deviation of population activity temporal dynamics
	($\hat{\sigma}/\hat{\sigma}_0$) as a function of $\rho$ (averaged over
	20 independent trials with an expected number of 650 updates per unit).
	Simulations for the \textbf{inset} were performed using the stochastic
	update scheme described in \cite{van_vreeswijk_chaotic_1998}.
	Parameters: $N=5000$, $\bar{J}=-0.7$ and all other parameters are as in
	Fig.\ref{fig1}.} \label{fig2} \end{figure}

In the second scenario, we show that an inherently stochastic mean-field limit
with nontrivial fluctuations can emerge in a network with statistically
inhomogeneous out-degrees.
The condition in eqn.~\ref{CI} guarantees the convergence of the average
population activity to the prediction of MFT. Indeed eqn.~\ref{CI} indicates
that, as $N\rightarrow\infty$, the average column sum of the connectivity
matrix $\mathbf{J}$ should be $K$. It is straightforward to construct networks
that do not obey this rule; such networks lose their pointwise convergence to a
deterministic MFT in eqn.~\ref{MFE}. 
An extreme example of a network of this kind is a network that has a single
unit, $n_{j_*}$, that connects into $\rho N$ units in the circuits, where
$0<\rho \leq 1$ is the fraction of units in the network that are post-synaptic
for $n_{j_*}$.  Numerical simulations of such a network ($N=5000$ and $\rho=1$)
show large-amplitude population activity fluctuations (Fig.\ref{fig2}, black
line), in contrast to the smaller fluctuations of a homogeneous network
(Fig.\ref{fig2}, gray line). 
Our approach allows the construction of stochastic correction terms to the
mean-field limit by isolating the unit $n_{j_*}$ from the network and then
taking the limit $N \rightarrow \infty$. Therefore, a first-order correction to
the function $F$ of eqn.~\ref{F} can be derived as
\begin{equation} F_s(m(t)) \approx F(m(t)) + \rho
\bar{J}K^{-\gamma}f'(\mu_1(t)) n_{j_*}(t).  \label{F_s} \end{equation}
$F_s$ is a stochastic function since $n_{j_*}$ is a binomial random variable
for which the probability of being at state one is $m(t)$; the mean-field
equation is thus transformed into an ordinary stochastic differential equation.
The correction term in eqn.~\ref{F_s} indicates that the observed large
fluctuations (Fig.\ref{fig2}, green line) are indeed a finite $K$ phenomenon.
Therefore, in large networks that have a finite number of connections between
units (e.g. finite $K$ networks), it suffices that only one unit breaks the
condition (i.e.,  $\rho>0$) and, as a result, the deterministic MFT collapses
(Fig.\ref{fig2}, inset). The emergence of large-amplitude population events in
Fig.\ref{fig2} has been observed previously as the indication of the  ``synfire
chain'' in cortical networks simulations \cite{aviel_embedding_2003}.  It is
noteworthy that there is compelling evidence that a few neurons can form an
extensive number of post-synaptic connections in cortical microcircuits
\cite{lee_anatomy_2016}.  
In eqn.~\ref{F_s},  we observe that a unit with a high out-degree can influence
the macroscopic dynamics of the system.  Therefore, recent experimental results
that indicate the diverse couplings between single-cell activity and population
averages in cortical networks \cite{okun_diverse_2015} can be the result of
inhomogeneity of out-degrees.


In this {\it letter}, we studied a simplified model that captures the essential
nonequilibrium aspects of cortical asynchronous state
\cite{van_vreeswijk_chaotic_1998}, and it allowed us to demonstrate the
calculation of a complete statistics of fluctuations in fixed indegree
networks. Our results show that the MFT for
binary units can be fundamentally constructed from the
LLNs and the emergence of intrinsic fluctuations does not require the
application of the central limit theorem. 
It is noteworthy that considering other heterogeneities in the system requires
extra averaging operations and performing self-consistent calculations of
temporal and spatial fluctuations. For instance, in a network in which unit $i$
has $K_i$ incoming connections, it can be shown that the power of $m$ in the
eqn.~\ref{mu} must be replaced by moments of the rate distribution,
$\operatorname{E}\big(m^r\big)$, which can be self-consistently
determined.

The semimartingale decomposition captures the finite-size effect
(eqn.~\ref{OU_rate}) in the orthogonal direction to the average correlations
between units. These correlations have been investigated previously
\cite{renart_asynchronous_2010}. 
Here, the finite-size scaling of fluctuations can be derived directly form rate
matrix $Q$. 
In a recent study by Dahmen et al. \cite{dahmen_correlated_2016} the MFT of
binary units is extended by a cumulant expansion that allows the systematic
calculation of finite-size corrections to cumulants of arbitrary order.
In contrast, in our analysis all remaining correlations are implicitly
encapsulated in the martingale part.
Importantly, the application of martingale theory and the expansion of the
averaging operator allows a tractable alternative to the perturbative expansion
of the system's state-space evolution to formulate an exact theory for network
collective dynamics.

Our approach in this letter goes beyond the classical asymptotic analysis of
random connectivities, which requires statistical conditions for connectivity matrices,
$\mathbf{J}$, to ensure pointwise convergence to a deterministic MFT,
irrespective of any fine or major motifs in the connectivity matrix, and
suggests a universality class of MFT for all fixed-in-degree and {\it iid}
networks.  Furthermore, we demonstrated a computationally interesting
phenomenon for emergence of a stochastic MFT by breaking the first condition
(eqn.~\ref{CI}).  
Our framework can be readily exploited to determine the mean-field equilibrium
of symmetrically disordered systems, in the presence of microstructures in
their interactions such as spin glasses and associative neural networks \cite{
	agliari_multitasking_2012, *sollich_extensive_2014}, similarly.
Once the connectivity matrix is given, it is straightforward to determine if
the system's mean population activity converges to the MFT in an annealed
dynamics with independent initial conditions.  In the quenched dynamics, the
analysis of metastability requires a further investigation of the invariance
measures of the state-space. Taken all together, we believe this approach paves
the way for investigating the MFT of various networks collective
phenomena.


{\it Acknowledgment}. This work was supported by the Federal Ministry of
Education and Research (BMBF) Germany under grant No.~FKZ01GQ1001B and Deutsche
Forschungsgemeinschaft (DFG) grant No.~FA1316/2-1.


\bibliography{prl_ref}

\begin{thebibliography}{24}%
\makeatletter
\providecommand \@ifxundefined [1]{%
 \@ifx{#1\undefined}
}%
\providecommand \@ifnum [1]{%
 \ifnum #1\expandafter \@firstoftwo
 \else \expandafter \@secondoftwo
 \fi
}%
\providecommand \@ifx [1]{%
 \ifx #1\expandafter \@firstoftwo
 \else \expandafter \@secondoftwo
 \fi
}%
\providecommand \natexlab [1]{#1}%
\providecommand \enquote  [1]{``#1''}%
\providecommand \bibnamefont  [1]{#1}%
\providecommand \bibfnamefont [1]{#1}%
\providecommand \citenamefont [1]{#1}%
\providecommand \href@noop [0]{\@secondoftwo}%
\providecommand \href [0]{\begingroup \@sanitize@url \@href}%
\providecommand \@href[1]{\@@startlink{#1}\@@href}%
\providecommand \@@href[1]{\endgroup#1\@@endlink}%
\providecommand \@sanitize@url [0]{\catcode `\\12\catcode `\$12\catcode
  `\&12\catcode `\#12\catcode `\^12\catcode `\_12\catcode `\%12\relax}%
\providecommand \@@startlink[1]{}%
\providecommand \@@endlink[0]{}%
\providecommand \url  [0]{\begingroup\@sanitize@url \@url }%
\providecommand \@url [1]{\endgroup\@href {#1}{\urlprefix }}%
\providecommand \urlprefix  [0]{URL }%
\providecommand \Eprint [0]{\href }%
\providecommand \doibase [0]{http://dx.doi.org/}%
\providecommand \selectlanguage [0]{\@gobble}%
\providecommand \bibinfo  [0]{\@secondoftwo}%
\providecommand \bibfield  [0]{\@secondoftwo}%
\providecommand \translation [1]{[#1]}%
\providecommand \BibitemOpen [0]{}%
\providecommand \bibitemStop [0]{}%
\providecommand \bibitemNoStop [0]{.\EOS\space}%
\providecommand \EOS [0]{\spacefactor3000\relax}%
\providecommand \BibitemShut  [1]{\csname bibitem#1\endcsname}%
\let\auto@bib@innerbib\@empty
\bibitem [{\citenamefont {Mézard}\ \emph {et~al.}(1986)\citenamefont
  {Mézard}, \citenamefont {Parisi},\ and\ \citenamefont
  {Virasoro}}]{mezard_spin_1986}%
  \BibitemOpen
  \bibfield  {author} {\bibinfo {author} {\bibfnamefont {M.}~\bibnamefont
  {Mézard}}, \bibinfo {author} {\bibfnamefont {G.}~\bibnamefont {Parisi}}, \
  and\ \bibinfo {author} {\bibfnamefont {M.}~\bibnamefont {Virasoro}},\ }\href
  {\doibase 04556 DOI: 10.1142/0271} {\emph {\bibinfo {title} {Spin {Glass}
  {Theory} and {Beyond}: {An} {Introduction} to the {Replica} {Method} and
  {Its} {Applications}}}},\ \bibinfo {series} {World {Scientific} {Lecture}
  {Notes} in {Physics}}, Vol.~\bibinfo {volume} {9}\ (\bibinfo  {publisher}
  {World Scientific},\ \bibinfo {year} {1986})\BibitemShut {NoStop}%
\bibitem [{\citenamefont {Ising}(1925)}]{ising_beitrag_1925}%
  \BibitemOpen
  \bibfield  {author} {\bibinfo {author} {\bibfnamefont {E.}~\bibnamefont
  {Ising}},\ }\href {\doibase 10.1007/BF02980577} {\bibfield  {journal}
  {\bibinfo  {journal} {Zeitschrift für Physik}\ }\textbf {\bibinfo {volume}
  {31}},\ \bibinfo {pages} {253} (\bibinfo {year} {1925})}\BibitemShut
  {NoStop}%
\bibitem [{\citenamefont {Hohenberg}\ and\ \citenamefont
  {Halperin}(1977)}]{hohenberg_theory_1977}%
  \BibitemOpen
  \bibfield  {author} {\bibinfo {author} {\bibfnamefont {P.~C.}\ \bibnamefont
  {Hohenberg}}\ and\ \bibinfo {author} {\bibfnamefont {B.~I.}\ \bibnamefont
  {Halperin}},\ }\href {\doibase 10.1103/RevModPhys.49.435} {\bibfield
  {journal} {\bibinfo  {journal} {Reviews of Modern Physics}\ }\textbf
  {\bibinfo {volume} {49}},\ \bibinfo {pages} {435} (\bibinfo {year}
  {1977})}\BibitemShut {NoStop}%
\bibitem [{\citenamefont {Hamley}(2013)}]{hamley_introduction_2013}%
  \BibitemOpen
  \bibfield  {author} {\bibinfo {author} {\bibfnamefont {I.~W.}\ \bibnamefont
  {Hamley}},\ }\href@noop {} {\emph {\bibinfo {title} {Introduction to {Soft}
  {Matter}: {Synthetic} and {Biological} {Self}-{Assembling} {Materials}}}}\
  (\bibinfo  {publisher} {John Wiley \& Sons},\ \bibinfo {year}
  {2013})\BibitemShut {NoStop}%
\bibitem [{\citenamefont {van Vreeswijk}\ and\ \citenamefont
  {Sompolinsky}(1998)}]{van_vreeswijk_chaotic_1998}%
  \BibitemOpen
  \bibfield  {author} {\bibinfo {author} {\bibfnamefont {C.}~\bibnamefont {van
  Vreeswijk}}\ and\ \bibinfo {author} {\bibfnamefont {H.}~\bibnamefont
  {Sompolinsky}},\ }\href@noop {} {\bibfield  {journal} {\bibinfo  {journal}
  {Neural Comput}\ }\textbf {\bibinfo {volume} {10}},\ \bibinfo {pages} {1321}
  (\bibinfo {year} {1998})}\BibitemShut {NoStop}%
\bibitem [{\citenamefont {Glauber}(1963)}]{glauber_timedependent_1963}%
  \BibitemOpen
  \bibfield  {author} {\bibinfo {author} {\bibfnamefont {R.~J.}\ \bibnamefont
  {Glauber}},\ }\href {\doibase 10.1063/1.1703954} {\bibfield  {journal}
  {\bibinfo  {journal} {Journal of Mathematical Physics}\ }\textbf {\bibinfo
  {volume} {4}},\ \bibinfo {pages} {294} (\bibinfo {year} {1963})}\BibitemShut
  {NoStop}%
\bibitem [{\citenamefont {Ginzburg}\ and\ \citenamefont
  {Sompolinsky}(1994)}]{ginzburg_theory_1994}%
  \BibitemOpen
  \bibfield  {author} {\bibinfo {author} {\bibfnamefont {I.}~\bibnamefont
  {Ginzburg}}\ and\ \bibinfo {author} {\bibfnamefont {H.}~\bibnamefont
  {Sompolinsky}},\ }\href {\doibase 10.1103/PhysRevE.50.3171} {\bibfield
  {journal} {\bibinfo  {journal} {Physical Review E}\ }\textbf {\bibinfo
  {volume} {50}},\ \bibinfo {pages} {3171} (\bibinfo {year}
  {1994})}\BibitemShut {NoStop}%
\bibitem [{\citenamefont {Buice}\ \emph {et~al.}(2010)\citenamefont {Buice},
  \citenamefont {Cowan},\ and\ \citenamefont {Chow}}]{buice_systematic_2010}%
  \BibitemOpen
  \bibfield  {author} {\bibinfo {author} {\bibfnamefont {M.~A.}\ \bibnamefont
  {Buice}}, \bibinfo {author} {\bibfnamefont {J.~D.}\ \bibnamefont {Cowan}}, \
  and\ \bibinfo {author} {\bibfnamefont {C.~C.}\ \bibnamefont {Chow}},\ }\href
  {\doibase 10.1162/neco.2009.02-09-960} {\bibfield  {journal} {\bibinfo
  {journal} {Neural computation}\ }\textbf {\bibinfo {volume} {22}},\ \bibinfo
  {pages} {377} (\bibinfo {year} {2010})}\BibitemShut {NoStop}%
\bibitem [{\citenamefont {Renart}\ \emph {et~al.}(2010)\citenamefont {Renart},
  \citenamefont {Rocha}, \citenamefont {Bartho}, \citenamefont {Hollender},
  \citenamefont {Parga}, \citenamefont {Reyes},\ and\ \citenamefont
  {Harris}}]{renart_asynchronous_2010}%
  \BibitemOpen
  \bibfield  {author} {\bibinfo {author} {\bibfnamefont {A.}~\bibnamefont
  {Renart}}, \bibinfo {author} {\bibfnamefont {J.~d.~l.}\ \bibnamefont
  {Rocha}}, \bibinfo {author} {\bibfnamefont {P.}~\bibnamefont {Bartho}},
  \bibinfo {author} {\bibfnamefont {L.}~\bibnamefont {Hollender}}, \bibinfo
  {author} {\bibfnamefont {N.}~\bibnamefont {Parga}}, \bibinfo {author}
  {\bibfnamefont {A.}~\bibnamefont {Reyes}}, \ and\ \bibinfo {author}
  {\bibfnamefont {K.~D.}\ \bibnamefont {Harris}},\ }\href {\doibase
  10.1126/science.1179850} {\bibfield  {journal} {\bibinfo  {journal}
  {Science}\ }\textbf {\bibinfo {volume} {327}},\ \bibinfo {pages} {587}
  (\bibinfo {year} {2010})}\BibitemShut {NoStop}%
\bibitem [{\citenamefont {Mézard}\ and\ \citenamefont
  {Sakellariou}(2011)}]{mezard_exact_2011}%
  \BibitemOpen
  \bibfield  {author} {\bibinfo {author} {\bibfnamefont {M.}~\bibnamefont
  {Mézard}}\ and\ \bibinfo {author} {\bibfnamefont {J.}~\bibnamefont
  {Sakellariou}},\ }\href {\doibase 10.1088/1742-5468/2011/07/L07001}
  {\bibfield  {journal} {\bibinfo  {journal} {Journal of Statistical Mechanics:
  Theory and Experiment}\ }\textbf {\bibinfo {volume} {2011}},\ \bibinfo
  {pages} {L07001} (\bibinfo {year} {2011})}\BibitemShut {NoStop}%
\bibitem [{\citenamefont {Stroock}(2008)}]{stroock_introduction_2008}%
  \BibitemOpen
  \bibfield  {author} {\bibinfo {author} {\bibfnamefont {D.~W.}\ \bibnamefont
  {Stroock}},\ }\href@noop {} {\emph {\bibinfo {title} {An {Introduction} to
  {Markov} {Processes}}}},\ \bibinfo {edition} {2005th}\ ed.\ (\bibinfo
  {publisher} {Springer},\ \bibinfo {address} {Berlin; New York},\ \bibinfo
  {year} {2008})\BibitemShut {NoStop}%
\bibitem [{sup()}]{supp}%
  \BibitemOpen
  \href {http://link.aps.org/doi} {\ }\BibitemShut {NoStop}%
\bibitem [{\citenamefont {Gillespie}(1976)}]{gillespie_general_1976}%
  \BibitemOpen
  \bibfield  {author} {\bibinfo {author} {\bibfnamefont {D.~T.}\ \bibnamefont
  {Gillespie}},\ }\href {\doibase 10.1016/0021-9991(76)90041-3} {\bibfield
  {journal} {\bibinfo  {journal} {Journal of Computational Physics}\ }\textbf
  {\bibinfo {volume} {22}},\ \bibinfo {pages} {403} (\bibinfo {year}
  {1976})}\BibitemShut {NoStop}%
\bibitem [{\citenamefont {Doob}(1942)}]{doob_topics_1942}%
  \BibitemOpen
  \bibfield  {author} {\bibinfo {author} {\bibfnamefont {J.~L.}\ \bibnamefont
  {Doob}},\ }\href {\doibase 10.1090/S0002-9947-1942-0006633-7} {\bibfield
  {journal} {\bibinfo  {journal} {Transactions of the American Mathematical
  Society}\ }\textbf {\bibinfo {volume} {52}},\ \bibinfo {pages} {37} (\bibinfo
  {year} {1942})}\BibitemShut {NoStop}%
\bibitem [{\citenamefont {Dahmen}\ \emph {et~al.}(2016)\citenamefont {Dahmen},
  \citenamefont {Bos},\ and\ \citenamefont {Helias}}]{dahmen_correlated_2016}%
  \BibitemOpen
  \bibfield  {author} {\bibinfo {author} {\bibfnamefont {D.}~\bibnamefont
  {Dahmen}}, \bibinfo {author} {\bibfnamefont {H.}~\bibnamefont {Bos}}, \ and\
  \bibinfo {author} {\bibfnamefont {M.}~\bibnamefont {Helias}},\ }\href
  {\doibase 10.1103/PhysRevX.6.031024} {\bibfield  {journal} {\bibinfo
  {journal} {Physical Review X}\ }\textbf {\bibinfo {volume} {6}},\ \bibinfo
  {pages} {031024} (\bibinfo {year} {2016})}\BibitemShut {NoStop}%
\bibitem [{\citenamefont {Stuard}\ and\ \citenamefont
  {Ord}(1994)}]{stuard_kendalls_1994}%
  \BibitemOpen
  \bibfield  {author} {\bibinfo {author} {\bibfnamefont {A.}~\bibnamefont
  {Stuard}}\ and\ \bibinfo {author} {\bibfnamefont {J.~K.}\ \bibnamefont
  {Ord}},\ }\href@noop {} {\emph {\bibinfo {title} {Kendall’s {Advanced}
  {Theory} of {Statistics}. {Vol}. 1. {Distribution} {Theory}}}}\ (\bibinfo
  {publisher} {Halsted Press, New York, NY},\ \bibinfo {year}
  {1994})\BibitemShut {NoStop}%
\bibitem [{\citenamefont {Brunel}\ and\ \citenamefont
  {Hakim}(1999)}]{brunel_fast_1999}%
  \BibitemOpen
  \bibfield  {author} {\bibinfo {author} {\bibfnamefont {N.}~\bibnamefont
  {Brunel}}\ and\ \bibinfo {author} {\bibfnamefont {V.}~\bibnamefont {Hakim}},\
  }\href {\doibase 10.1162/089976699300016179} {\bibfield  {journal} {\bibinfo
  {journal} {Neural Computation}\ }\textbf {\bibinfo {volume} {11}},\ \bibinfo
  {pages} {1621} (\bibinfo {year} {1999})}\BibitemShut {NoStop}%
\bibitem [{\citenamefont {Mattia}\ and\ \citenamefont
  {Giudice}(2002)}]{mattia_mean-field_2002}%
  \BibitemOpen
  \bibfield  {author} {\bibinfo {author} {\bibfnamefont {M.}~\bibnamefont
  {Mattia}}\ and\ \bibinfo {author} {\bibfnamefont {P.~D.}\ \bibnamefont
  {Giudice}},\ }in\ \href@noop {} {\emph {\bibinfo {booktitle} {Artificial
  {Neural} {Networks} — {ICANN} 2002}}},\ \bibinfo {series and number}
  {\bibinfo {series} {Lecture {Notes} in {Computer} {Science}}\ No.\ \bibinfo
  {number} {2415}},\ \bibinfo {editor} {edited by\ \bibinfo {editor}
  {\bibfnamefont {J.~R.}\ \bibnamefont {Dorronsoro}}}\ (\bibinfo  {publisher}
  {Springer Berlin Heidelberg},\ \bibinfo {year} {2002})\ pp.\ \bibinfo {pages}
  {111--116}\BibitemShut {NoStop}%
\bibitem [{\citenamefont {Grytskyy}\ \emph {et~al.}(2013)\citenamefont
  {Grytskyy}, \citenamefont {Tetzlaff}, \citenamefont {Diesmann},\ and\
  \citenamefont {Helias}}]{grytskyy_unified_2013}%
  \BibitemOpen
  \bibfield  {author} {\bibinfo {author} {\bibfnamefont {D.}~\bibnamefont
  {Grytskyy}}, \bibinfo {author} {\bibfnamefont {T.}~\bibnamefont {Tetzlaff}},
  \bibinfo {author} {\bibfnamefont {M.}~\bibnamefont {Diesmann}}, \ and\
  \bibinfo {author} {\bibfnamefont {M.}~\bibnamefont {Helias}},\ }\href
  {\doibase 10.3389/fncom.2013.00131} {\bibfield  {journal} {\bibinfo
  {journal} {Frontiers in Computational Neuroscience}\ }\textbf {\bibinfo
  {volume} {7}} (\bibinfo {year} {2013}),\
  10.3389/fncom.2013.00131}\BibitemShut {NoStop}%
\bibitem [{\citenamefont {Aviel}\ \emph {et~al.}(2003)\citenamefont {Aviel},
  \citenamefont {Mehring}, \citenamefont {Abeles},\ and\ \citenamefont
  {Horn}}]{aviel_embedding_2003}%
  \BibitemOpen
  \bibfield  {author} {\bibinfo {author} {\bibfnamefont {Y.}~\bibnamefont
  {Aviel}}, \bibinfo {author} {\bibfnamefont {C.}~\bibnamefont {Mehring}},
  \bibinfo {author} {\bibfnamefont {M.}~\bibnamefont {Abeles}}, \ and\ \bibinfo
  {author} {\bibfnamefont {D.}~\bibnamefont {Horn}},\ }\href
  {http://www.mitpressjournals.org/doi/abs/10.1162/089976603321780290}
  {\bibfield  {journal} {\bibinfo  {journal} {Neural computation}\ }\textbf
  {\bibinfo {volume} {15}},\ \bibinfo {pages} {1321} (\bibinfo {year}
  {2003})}\BibitemShut {NoStop}%
\bibitem [{\citenamefont {Lee}\ \emph {et~al.}(2016)\citenamefont {Lee},
  \citenamefont {Bonin}, \citenamefont {Reed}, \citenamefont {Graham},
  \citenamefont {Hood}, \citenamefont {Glattfelder},\ and\ \citenamefont
  {Reid}}]{lee_anatomy_2016}%
  \BibitemOpen
  \bibfield  {author} {\bibinfo {author} {\bibfnamefont {W.-C.~A.}\
  \bibnamefont {Lee}}, \bibinfo {author} {\bibfnamefont {V.}~\bibnamefont
  {Bonin}}, \bibinfo {author} {\bibfnamefont {M.}~\bibnamefont {Reed}},
  \bibinfo {author} {\bibfnamefont {B.~J.}\ \bibnamefont {Graham}}, \bibinfo
  {author} {\bibfnamefont {G.}~\bibnamefont {Hood}}, \bibinfo {author}
  {\bibfnamefont {K.}~\bibnamefont {Glattfelder}}, \ and\ \bibinfo {author}
  {\bibfnamefont {R.~C.}\ \bibnamefont {Reid}},\ }\href {\doibase
  10.1038/nature17192} {\bibfield  {journal} {\bibinfo  {journal} {Nature}\
  }\textbf {\bibinfo {volume} {532}},\ \bibinfo {pages} {370} (\bibinfo {year}
  {2016})}\BibitemShut {NoStop}%
\bibitem [{\citenamefont {Okun}\ \emph {et~al.}(2015)\citenamefont {Okun},
  \citenamefont {Steinmetz}, \citenamefont {Cossell}, \citenamefont {Iacaruso},
  \citenamefont {Ko}, \citenamefont {Barthó}, \citenamefont {Moore},
  \citenamefont {Hofer}, \citenamefont {Mrsic-Flogel}, \citenamefont
  {Carandini},\ and\ \citenamefont {Harris}}]{okun_diverse_2015}%
  \BibitemOpen
  \bibfield  {author} {\bibinfo {author} {\bibfnamefont {M.}~\bibnamefont
  {Okun}}, \bibinfo {author} {\bibfnamefont {N.~A.}\ \bibnamefont {Steinmetz}},
  \bibinfo {author} {\bibfnamefont {L.}~\bibnamefont {Cossell}}, \bibinfo
  {author} {\bibfnamefont {M.~F.}\ \bibnamefont {Iacaruso}}, \bibinfo {author}
  {\bibfnamefont {H.}~\bibnamefont {Ko}}, \bibinfo {author} {\bibfnamefont
  {P.}~\bibnamefont {Barthó}}, \bibinfo {author} {\bibfnamefont
  {T.}~\bibnamefont {Moore}}, \bibinfo {author} {\bibfnamefont {S.~B.}\
  \bibnamefont {Hofer}}, \bibinfo {author} {\bibfnamefont {T.~D.}\ \bibnamefont
  {Mrsic-Flogel}}, \bibinfo {author} {\bibfnamefont {M.}~\bibnamefont
  {Carandini}}, \ and\ \bibinfo {author} {\bibfnamefont {K.~D.}\ \bibnamefont
  {Harris}},\ }\href {\doibase 10.1038/nature14273} {\bibfield  {journal}
  {\bibinfo  {journal} {Nature}\ }\textbf {\bibinfo {volume} {521}},\ \bibinfo
  {pages} {511} (\bibinfo {year} {2015})}\BibitemShut {NoStop}%
\bibitem [{\citenamefont {Agliari}\ \emph {et~al.}(2012)\citenamefont
  {Agliari}, \citenamefont {Barra}, \citenamefont {Galluzzi}, \citenamefont
  {Guerra},\ and\ \citenamefont {Moauro}}]{agliari_multitasking_2012}%
  \BibitemOpen
  \bibfield  {author} {\bibinfo {author} {\bibfnamefont {E.}~\bibnamefont
  {Agliari}}, \bibinfo {author} {\bibfnamefont {A.}~\bibnamefont {Barra}},
  \bibinfo {author} {\bibfnamefont {A.}~\bibnamefont {Galluzzi}}, \bibinfo
  {author} {\bibfnamefont {F.}~\bibnamefont {Guerra}}, \ and\ \bibinfo {author}
  {\bibfnamefont {F.}~\bibnamefont {Moauro}},\ }\href {\doibase
  10.1103/PhysRevLett.109.268101} {\bibfield  {journal} {\bibinfo  {journal}
  {Physical Review Letters}\ }\textbf {\bibinfo {volume} {109}} (\bibinfo
  {year} {2012}),\ 10.1103/PhysRevLett.109.268101}\BibitemShut {NoStop}%
\bibitem [{\citenamefont {Sollich}\ \emph {et~al.}(2014)\citenamefont
  {Sollich}, \citenamefont {Tantari}, \citenamefont {Annibale},\ and\
  \citenamefont {Barra}}]{sollich_extensive_2014}%
  \BibitemOpen
  \bibfield  {author} {\bibinfo {author} {\bibfnamefont {P.}~\bibnamefont
  {Sollich}}, \bibinfo {author} {\bibfnamefont {D.}~\bibnamefont {Tantari}},
  \bibinfo {author} {\bibfnamefont {A.}~\bibnamefont {Annibale}}, \ and\
  \bibinfo {author} {\bibfnamefont {A.}~\bibnamefont {Barra}},\ }\href
  {\doibase 10.1103/PhysRevLett.113.238106} {\bibfield  {journal} {\bibinfo
  {journal} {Physical Review Letters}\ }\textbf {\bibinfo {volume} {113}}
  (\bibinfo {year} {2014}),\ 10.1103/PhysRevLett.113.238106}\BibitemShut
  {NoStop}%
\end{thebibliography}%

\end{document}